\documentclass[runningheads]{llncs}

\usepackage{nicefrac}
\usepackage{siunitx}
\usepackage{array,framed}
\usepackage{booktabs}
\usepackage{
  color,
  float,
  epsfig,
  wrapfig,
  graphics,
  graphicx,
  subcaption,
}

\usepackage{caption,textcomp,amssymb}

\usepackage{latexsym,fancyhdr,url}
\usepackage{enumerate}
\graphicspath{ {images/} }
\usepackage{multirow}
\usepackage{balance}

\usepackage{mathtools}

\DeclareMathAlphabet{\mathcal}{OMS}{cmsy}{m}{n}

\usepackage{soul}
\soulregister\cite7
\soulregister\ac7
\soulregister\ref7
\graphicspath{ {images/} }

\usepackage{hyperref}

\usepackage{orcidlink}
\usepackage{acronym}

\newacro{IDS}[IDS]{Intrusion Detection System}
\newacro{DNN}[DNN]{Deep Neural Network}
\newacro{DBN}[DBN]{Deep Belief Network}
\newacro{MLP}[MLP]{Multi-Layer Perceptron}
\newacro{OT}[OT]{operational technology}
\newacro{DBN}[DBN]{Deep Belief Network}
\newacro{NIDS}[NIDS]{Network Intrusion Detection System}
\newacro{HIDS}[HIDS]{Host Intrusion Detection System}
\newacro{RBM}[RBM]{Restricted Boltzmann Machine}
\newacro{AI}[AI]{Artificial Intelligence}
\newacro{SVM}[SVM]{Support Vector Machine}
\newacro{RNN}[RNN]{Recurrent Neural Network}
\newacro{CNN}[CNN]{Convolutional Neural Network}
\newacro{DoS}[DoS]{Denial Of Service}
\newacro{DDoS}[DDoS]{Distributed Denial Of Service}
\newacro{PCA}[PCA]{Principal Components Analysis}
\newacro{SMOTE}[SMOTE]{Synthetic Minority Over-sampling Technique}

\begin{document}

\title{An Intrusion Detection System based on Deep Belief Networks}

\author{Othmane Belarbi\orcidlink{0000-0002-6106-7669}\inst{1} \and Aftab Khan\orcidlink{0000-0002-3573-6240}*\inst{1} \and Pietro Carnelli\orcidlink{0000-0002-4993-5873}\inst{1} \and Theodoros Spyridopoulos\orcidlink{0000-0001-7575-9909}\inst{2}}

\institute{Toshiba Europe Ltd., Bristol Research \& Innovation Laboratory, Bristol, UK \\ \email{\{othmane.belarbi, aftab.khan, pietro.carnelli\}@toshiba-bril.com}\\ \and
Cardiff University, School of Computer Science \& Informatics, Cardiff, UK \\ \email{spyridopoulost@cardiff.ac.uk}}

\authorrunning{Belarbi, Khan, Carnelli, Spyridopoulos}

\hyphenation{resampling}
\maketitle

%
%
\begin{abstract}
    The rapid growth of connected devices has led to the proliferation of novel cyber-security threats known as zero-day attacks. Traditional behaviour-based \acp{IDS} rely on \acp{DNN} to detect these attacks. The quality of the dataset used to train the \acp{DNN} plays a critical role in the detection performance, with underrepresented samples causing poor performances. In this paper, we develop and evaluate the performance of \acp{DBN} on detecting cyber-attacks within a network of connected devices. The CICIDS2017 dataset was used to train and evaluate the performance of our proposed \ac{DBN} approach. Several class balancing techniques were applied and evaluated. Lastly, we compare our approach against a conventional \ac{MLP} model and the existing state-of-the-art. Our proposed \ac{DBN} approach shows competitive and promising results, with significant performance improvement on the detection of attacks underrepresented in the training dataset.

    \keywords{Network Intrusion Detection System \and Deep Learning \and Deep Belief Networks \and Class Balancing}
\end{abstract}

%
%
\section{Introduction}
\label{sec:intro}

Traditional \acfp{IDS} are signature-based, relying on existing signatures of known attacks. They are designed to detect single attacks or families of attacks based on their particular characteristics. However, the vast amount of data that needs to be daily processed to develop new signatures renders their timely generation a challenging task. Furthermore, the generation of signatures for complex attacks that have evolved from previously known threats is not straightforward and requires additional effort \cite{Zhongw}.

Behaviour analysis techniques have also been explored by researchers to detect intrusions \cite{Zhongw}. Most of these systems use machine learning techniques to model system behaviour and detect malicious activity. Compared to signature-based \acp{IDS}, machine learning techniques can detect large families of attack variants regardless of their complexity. Typical machine learning techniques for classification are used to train models of malicious behaviour based on existing labelled datasets of system activity \cite{Alqatf,KUNANG2021102804}. Classification can be binary (i.e. normal or  malicious) or multi-class. In multi-class classification, the model can also detect the type of attack based on the classes/labels in the training set. 

The quality of the dataset plays a critical role in the detection performance of the trained \ac{IDS} model. When a type of attack is underrepresented in the dataset, typical in \ac{IDS} datasets, the resulting model performs poorly on the detection of attack variants that belong to the infrequent attack type. Several attempts have been proposed to mitigate the issues caused by imbalanced \ac{IDS} datasets, focusing mainly on the data sampling and class balancing techniques \cite{Sarpe}.

In this paper, we propose a multi-class classification \ac{NIDS} based on \acfp{DBN}\footnotemark{}. \ac{DBN} is a generative graphical model formed by stacking multiple \acp{RBM}. It can identify and learn high-dimensional representations. A \ac{DBN} is first pre-trained in an unsupervised way by using a greedy layer-by-layer learning algorithm  and then fine-tuned using the back-propagation technique in a supervised manner. As shown in our experimental results, this two-stage training process improves the detection performance against infrequent attack samples whilst retaining a high performance against the rest of the attacks.

This paper's contributions can be summarised as follows:
\begin{enumerate}
\item We demonstrate that our \ac{DBN}-based \ac{NIDS} outperforms \ac{MLP}-based \acp{NIDS}, especially when there is a small number of attack samples in the dataset.
\item We conducted a series of class balancing experiments on the highly imbalanced CICIDS2017 dataset \cite{Sharafaldin}, which includes benign network traffic and twelve attacks. Our experimental results demonstrate that our class balancing approach improves the detection performance in terms of F1-score.
\item The classification results of our \ac{NIDS}, after applying our class balancing approach, are compared against the state-of-the-art. Our proposed method demonstrates significant improvement in F1-score from 0.873 to 0.94.
\end{enumerate}

The rest of the paper is organised as follows. In Section \ref{sec:background and relwork}, we present recent related work on \acp{IDS} and \acp{DBN}. Section \ref{sec:methodology} describes the proposed methodology for the \ac{DBN}-based \ac{NIDS}. In Section \ref{sec:experiments}, we analyse the pre-processing approach we followed on the CICIDS2017 dataset and the model architecture we used, and report the performance results of the study comparing them against the state-of-the-art. Finally, we conclude the paper and present pathways for future work in Section \ref{sec:conclusion}.

\footnotetext{\url{https://github.com/othmbela/dbn-based-nids}}
\section{Related Work}
\label{sec:background and relwork}

In general, all ML-based \acp{IDS} comprise multiple components including data collection, pre-processing, feature extraction/selection and decision engine \cite{Bridgesra}. \ac{NIDS} rely on capturing and analysing inbound and outbound network traffic in the under analysis system. In most cases, they are separate devices attached to the network gateway. Depending on the application and the related privacy policy, they may collect and analyse only the headers of the network traffic packets or require access to the packet payload as well  \cite{Matyasv}. Our work focuses on the pre-processing, the feature extraction/selection and the decision engine for the implementation of \ac{NIDS}. Our \ac{NIDS} model relies on the CICIDS2017 dataset \cite{Sharafaldin}, which was generated by analysing full-packet network traffic. However, the extracted data and our generated features do not rely on the payload of the network packets nor on the source/destination IP address and port number. Therefore, our model preserves user privacy to an extent. Nevertheless, user privacy largely depends on the way data are captured on the network, which is outside the scope of our research. 

Various machine learning methodologies have been researched for the development of \ac{NIDS}~\cite{ASHFAQ2017484,GUO2016391,KIM20141690}. Among these, \acp{SVM} have been one of the most commonly used~\cite{CHITRAKAR2014231,GU2021102158,GU201953}. Researchers in~\cite{CHITRAKAR2014231} combined \acp{SVM} with ensemble learning to increase the generalisation ability of the model and improve its performance on the unknown data samples. However, both the Fuzzy-C Means (FCM) clustering method and the k-fold cross validation method used to train and test the final \ac{SVM} can be problematic when applied to network traffic time series datasets. A similar issue is evident in other current \ac{SVM}-based \ac{NIDS} works~\cite{GU2021102158}. Additionally, research conducted in~\cite{AnthiEirini} demonstrated that \acp{SVM} can take a significantly longer time to classify unseen data.

Recent work has demonstrated that deep learning techniques are quite efficient in identifying cyber-attacks on networks~\cite{Singlaan}. In~\cite{Yin}, the authors proposed a deep learning approach for intrusion detection using \acp{RNN}. The proposed \ac{RNN}-\ac{IDS} effectively recognised the type of intrusion with a higher accuracy and detection rate than traditional ML-based \ac{IDS} in both binary and multi-class classification. However, it performed poorly against the minority classes in multi-class classification, a common issue in cyber-security datasets. In a similar work~\cite{RHODE2018578}, it was demonstrated that an ensemble of \acp{RNN} can perform binary classification with 94\% accuracy within the first 5 seconds of execution. However, no discussion was provided by the authors around the performance of the approach against the minority classes of imbalanced datasets. 

In \cite{1D-CNN-LSTM}, the authors implemented and tested four classifiers for \ac{DDoS} attack detection using deep learning models such as MLP, 1D-\ac{CNN}, LSTM and 1D-\ac{CNN}+LSTM. The models were built based on the CICIDS2017 dataset~\cite{Sharafaldin}. The dataset was balanced by using the duplicating method. The experimental results demonstrated that the \ac{CNN}+LSTM performed better than the rest of the deep learning models achieving an accuracy of $97.16$\%.

The researchers in~\cite{LUCID} presented a deep learning detection system for \ac{DDoS} attacks, called LUCID. LUCID is a \ac{CNN}-based \ac{IDS} for binary classification of \ac{DDoS} attacks. The proposed approach can be used in online resource-constrained environments thanks to its ability to ensure low processing overhead and attack detection time. The evaluation results showed that the \ac{CNN}-based \ac{IDS} recognises \ac{DDoS} attack on the CICIDS2017 dataset with an accuracy of $99.7$\%.

The authors in \cite{Whisper} focused on real time detection of malicious traffic in large-scale high-throughput networks. The proposed approach, named Whisper, leverages frequency domain analysis to extract and analyse sequential information of the network traffic. Whisper can effectively detect 42 sophisticated attacks from the WIDE MAWI dataset in high throughput with at least $90\%$ detection accuracy. Even though our work does not focus on large-scale networks, combining it with frequency domain analysis is an interesting concept we would like to explore in future work. 

Researchers in~\cite{Yang} used an \ac{RBM} to extract fundamental features from the traffic data and then train an \ac{SVM} classifier. This approach was able to shorten the training time whilst preserving performance of the classifier. In a similar work~\cite{Peng}, a \ac{NIDS} based on deep learning was built using the KDD CUP'99 dataset. The authors used an \ac{RBM} to extract higher-level features from the input data followed by a back-propagation neural network to classify intrusions. The suggested model showed significant improvement in accuracy over traditional ML-based \ac{NIDS}. Finally, a system that classifies network intrusions using \acp{DBN} (a stack of \acp{RBM}) was implemented in~\cite{Alom}. In that work, the \ac{DBN}-\ac{IDS} achieved an accuracy of around $97.5$\% using $40$\% of the dataset in the training process. However, the evaluation metrics were only limited to accuracy, with no discussion around recall and precision. The proven ability of \acp{RBM} to improve the performance of existing classifiers, along with the lack of an in-depth analysis of \acp{DBN} as \ac{NIDS} and the limited work on tackling imbalanced cyber-security datasets, inspired the research presented in this paper.

\subsection{Restricted Boltzmann Machines}
\label{subsec:rbm}

\begin{figure}[t]
\begin{subfigure}[b]{0.49\textwidth}
    \centering
    \includegraphics[width=1\textwidth]{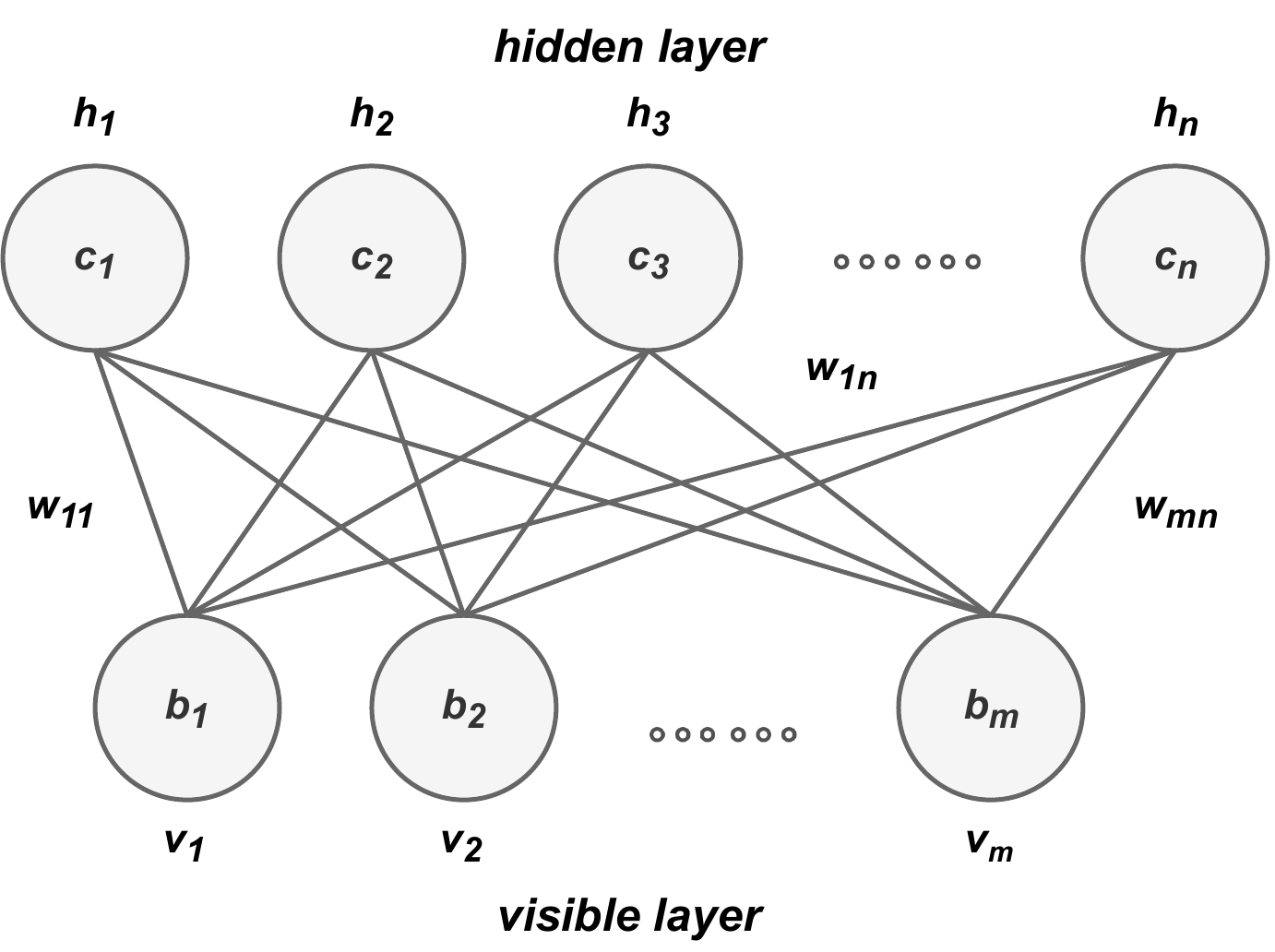}
    \caption{\ac{RBM}}
    \label{subfig:rbm}
\end{subfigure}
\hfill
\begin{subfigure}[b]{0.49\textwidth}
    \centering
    \includegraphics[width=1\textwidth]{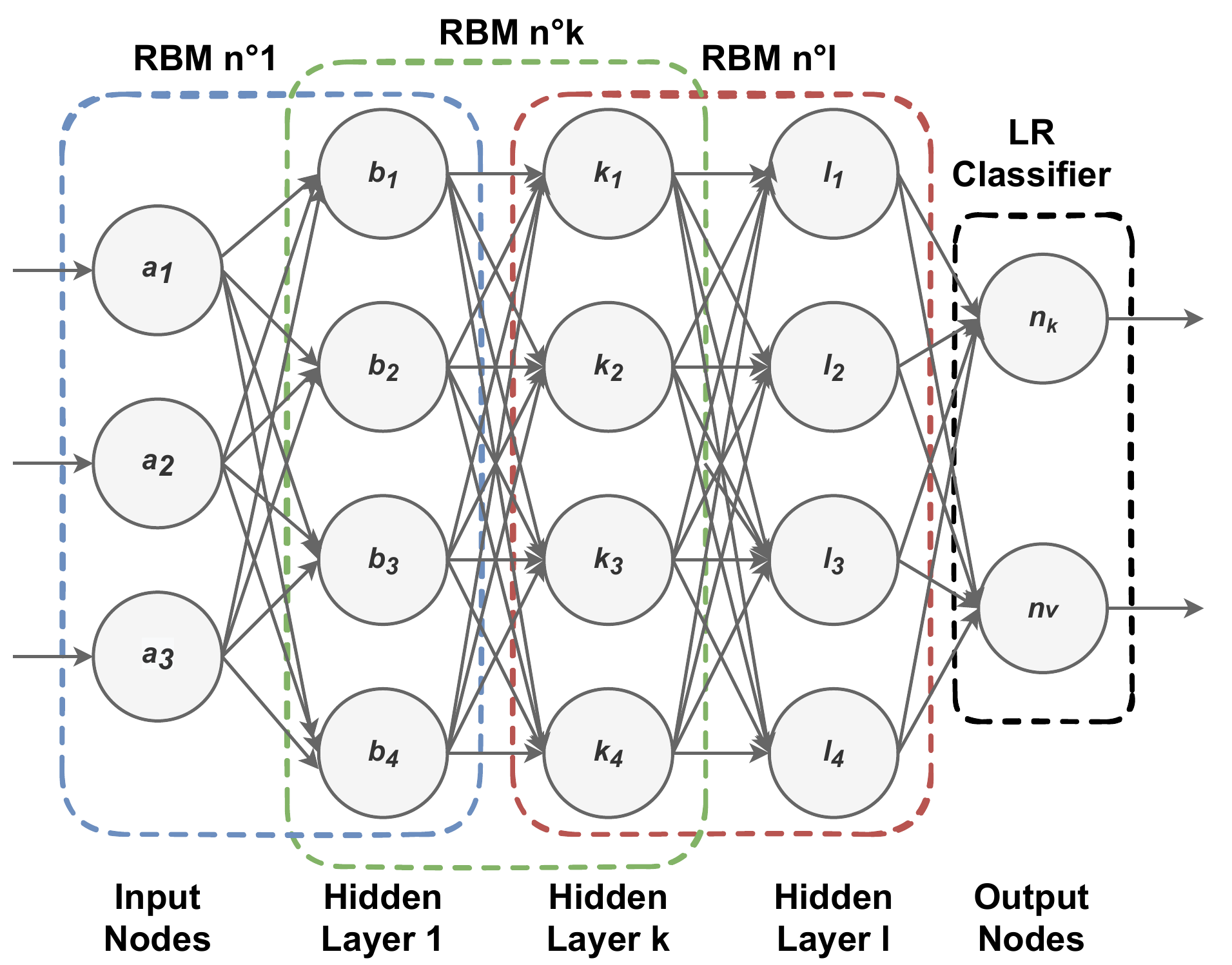}
    \caption{\ac{DBN}}
    \label{subfig:dbn}
\end{subfigure}
\caption{Algorithm structure}
\label{fig:algorithm structure}
\end{figure}

\acp{RBM} are Energy-Based Models (EBMs) that have been largely used for several tasks such as feature extraction, feature reduction and collaborative filtering \cite{Hinton,Salakhutdinov}. \acp{RBM} are two-layer undirected models where the layers are the visible layer and the hidden layer as illustrated in Fig. \ref{subfig:rbm}.


An \ac{RBM} captures the dependency between the visible and hidden units, $v$ and $h$, by associating an energy to each configuration of the variables. The energy function takes low values when the two variables are compatible and higher values when $h$ is less compatible with $v$.

The energy function of a joint configuration $(v, h)$ has an energy given by \cite{Geoffrey}:

\begin{equation}
E(v,h) = - \smashoperator{\sum_{i\in{visible}}} b_{i}v_{i} - \smashoperator{\sum_{j\in{hidden}}} c_{j}h_{j} - \smashoperator{\sum_{i,j}} v_{i}h_{j}w_{ij}
\end{equation}

where:
\begin{itemize}
     \item[--] $v_{i},h_{j}$ are the binary states of visible unit $i$ and hidden unit $j$.
     \item[--] $b_{i},c_{j}$ are the biases of visible unit $i$ and hidden unit $j$.
     \item[--] $w_{ij}$ is the weight between the visible unit $i$ and hidden unit $j$.
 \end{itemize}

The joint probability distribution of a pair of a visible vector $v$ and a hidden vector $h$ is defined via the energy function as:

\begin{equation}
p(v,h)=\frac{1}{Z}e^{-E(v,h)}
\end{equation}

where $Z$, the partition function, is defined as the summation over all possible pairs of visible and hidden vectors:

\begin{equation}
Z=\sum_{v, h}{e^{-E(v,h)}}
\end{equation}

The probability of the visible vector $v$ is given by marginalising out the hidden vector $h$:

\begin{equation}
p(v)=\frac{1}{Z}\sum_{h}{e^{-E(v,h)}}
\end{equation}

During the training phase, \acp{RBM} adjust their weights based on the Contrastive Divergence (CD) algorithm in which the second expectation term of the gradient descent, Equation \ref{equa:5}, is approximated after running k steps of the Gibbs sampler.

\begin{equation}
\Delta{w_{ij}}=\epsilon\big(<v_{i}h_{j}>_{data}-<v_{i}h_{j}>_{recon}\big)
\label{equa:5}
\end{equation}

where:
\begin{itemize}
 \item[--] $\epsilon$ is the learning rate.
 \item[--] $<v_{i}h_{j}>_{data}$ is the product of $v_{i}, h_{j}$ before reconstruction
 \item[--] $<v_{i}h_{j}>_{recon}$ is the expected value of $v_{i}, h_{j}$ after k-step reconstruction
\end{itemize}

\subsection{Deep Belief Network}
\label{subsec:dbn}

A \ac{DBN} is a deep architecture of multiple stacks of \acp{RBM} sequentially connected as shown in Fig. \ref{subfig:dbn}. Each \ac{RBM} model performs a non-linear transformation on its input vectors and produces outputs vectors that will serve as input for the next \ac{RBM} model in the sequence. Except for the first and final layers of the \ac{DBN}, every layer serves as a hidden layer to the nodes that come before and as a visible/input layer to the nodes that comes after. \acp{DBN} can be used in both unsupervised learning for generating images and in supervised learning for classification \cite{Salama}. \acp{DBN} are pre-trained layer by layer and fine-tuned using the back-propagation technique.
\section{Methodology}
\label{sec:methodology}

\begin{figure}[t]
    \centering
    \includegraphics[width=\textwidth]{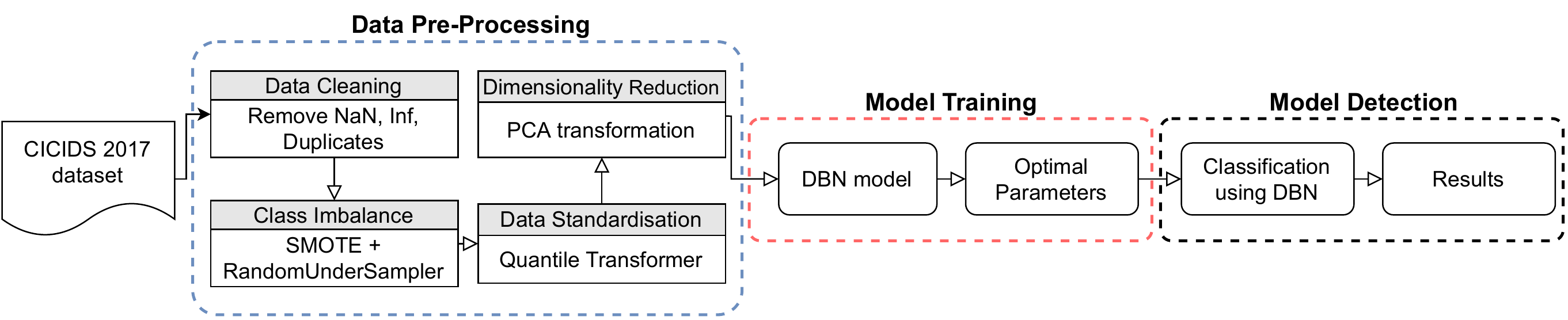}
    \caption{High-level architecture of our \ac{DBN}-based \ac{NIDS}}
    \label{fig:flowchart}
\end{figure}

In this paper, we aim to evaluate \acp{DBN} for network intrusion detection and address the issue of high-class imbalance in network traffic datasets.  Fig. \ref{fig:flowchart} illustrates the high level architecture of our approach. In the next sections we present and analyse the key processes in our architecture. 

\subsection{Training Data Pre-Processing}
\label{subsec:data processing}

We split the dataset into training, validation, and testing sets. The training set is used to train the model; the model learns the mapping function from this data. The validation set is used to perform the initial testing and tuning of the model, while the testing set is used to evaluate the model's performance.
In datasets with high-class imbalance (as is generally observed in \ac{IDS} datasets), a stratified split of the data is very important allowing a fairer assessment of the model's performance \cite{Hammerla}. 
Stratified data splitting preserves the proportion of samples for each target class essentially ensuring that enough samples of the minority class are included in the training set.

\textbf{\textit{Class Imbalance:}}
As highlighted above, \ac{IDS} datasets generally suffer from higher class imbalance compared to datasets from other domains.
For example, datasets such as NSL-KDD and CICIDS2017 have $95$\% and $90$\% benign samples respectively, with the rest containing different types of attacks.
This can fundamentally be attributed to the nature of these attacks, even in real-world deployments where the majority of the network traffic is expected to be benign in nature.
High-class imbalance in datasets introduces bias in favour of the majority class (benign), which makes the classification of the minority classes very challenging. There are multiple methods introduced in the literature that can handle such class imbalance. We evaluate the following methods in this paper: 

\textit{Under-sampling and Over-sampling:}
Under-sampling methods re-sample the \emph{majority} class by removing some of the instances.
Typically, a random under-sampler is employed in which the majority class samples are removed at random.
Over-sampling methods on the other hand re-sample the \emph{minority} classes by duplicating or creating new artificial instances with algorithms such as the \ac{SMOTE} \cite{SMOTE}.

\textit{Class Weight Strategy:}
Models could consider the underrepresented classes without re-sampling the training set by adding weights to the loss function of the algorithm \cite{ClassWeight}. The class weight strategy penalises the wrong classification of the minority classes more than the misclassification of the majority class. Hence, each class will have equal importance on gradient updates, on average, regardless of the number of samples. 

\textit{Sample Weight Strategy:}
Another method consists of associating a weight to each training sample where the weight is computed as described above. In this strategy, the dataset is re-balanced by ensuring that each batch of data is proportionally distributed.

\textbf{\textit{Data Standardisation:}}
\ac{IDS} datasets are sometimes high dimensional in nature (e.g. CICIDS2017 dataset has 78 features). In the case of such higher-dimensional datasets without any form of normalization, machine learning models in general do not optimise very well or take much longer to train. Standardising the input variables to overcome this issue can become skewed or biased, due to the large number of outliers in the highly imbalanced security datasets. Dropping outliers may help, however, it is not recommended since there is a risk of losing essential information required for correctly classifying attacks. There are different methods to overcome this problem by using statistics that are robust to outliers. Unlike other scalers, robust scaler and quantile transformation methods are not influenced by the impact of marginal outliers (because they are based on percentiles and quantiles respectively). In robust scaling, the median 
of a given feature is subtracted from values and divided by the interquartile range. The resulting range is larger than the other scalers. The quantile transformation is another standardization method that is robust to outliers. This method estimates the cumulative distribution function of the input variables and transforms the values to a uniform distribution [0, 1]. Then, the obtained values are mapped to the desired distribution using the associated quantile function.

\textbf{\textit{Dimensionality Reduction:}}
One of the most important steps in typical machine learning toolchains is dimensionality reduction. This can be achieved through feature selection or other methods such as \ac{PCA}. \ac{PCA} is a data analysis technique that transforms the possible correlated features set to uncorrelated features set. 
It can be used to reduce the dimensionality feature space by ignoring the components containing the least variance of the original feature space. 

\subsection{Model Training}
\label{subsec:model training}
The \ac{DBN} model requires two steps in the training process: unsupervised pre-training and supervised fine-tuning. In the first phase, each \ac{RBM} is trained to reconstruct its input by adjusting its weights and feeding the input layer of the next \ac{RBM}. This process is repeated until each \ac{RBM} layer is pre-trained (greedy learning). In supervised fine-tuning, all the weights are optimised by using the stochastic gradient descent and back-propagation.
Details of the specific architectures, hyperparameters, and performance metrics are provided in Section \ref{sec:experiments}. 
Finally, in the case of intrusion datasets that mainly suffer from high-class imbalance, the accuracy metric cannot be relied upon \cite{Galar}. Therefore, we chose the F1-score, precision, and recall as our metrics to evaluate the models.
\section{Experiments}
\label{sec:experiments}

All the experiments were conducted using a 64-bit Intel(R) Core(TM) i7-7500U CPU with 16GB RAM in Windows 10 environment. The models have been implemented in Python v3.8.2 using the PyTorch v1.9.0 library.

\subsection{Dataset}
\label{subsec:dataset}

\begin{table}[t]
\centering
\caption{Class distribution of CICIDS2017 dataset.}
\begin{tabular}{p{2.5cm}p{5cm}p{2cm}}
\toprule
\bf Category & \bf Labels & \bf \# samples \\
\midrule
Benign                         & Benign                          & 1,807,787                  \\\midrule
\multirow{3}{6em}{DoS/DDoS}    & Heartbleed, DDoS                & \multirow{3}{6em}{320,269} \\
                               & DoS Hulk, DoS GoldenEye         &                            \\
                               & DoS Slowloris, DoS Slowhttptest &                            \\\midrule
PortScan                       & PortScan                        & 57,305                     \\\midrule
Brute Force                    & FTP-Patator, SSH-Patator        & 8,551                      \\\midrule
\multirow{3}{6em}{Web Attack}  & Web Attack – Brute Force        & \multirow{3}{6em}{2,118}   \\
                               & Web Attack – XSS                &                            \\
                               & Web Attack – SQL Injection      &                            \\\midrule
Botnet                         & Bot                             & 1,943                      \\
\bottomrule
\end{tabular}
\label{tab:class_distribution}
\end{table}

In order to test the proposed \ac{DBN}-based NIDS, we used the CICIDS2017 dataset created by the Canadian Institute for Cyber-security (CIC) which satisfies the eleven criteria described in \cite{Sharafaldin}. The dataset includes samples of benign activity and common attacks over a period of five days. A B-Profile system described in \cite{Sharafaldin} was used to generate realistic benign traffic. The benign traffic corresponds to the human interaction of 25 users based on standard network protocols such as HTTP(S), FTP, SSH, IMAP and POP3. Several attacks were then initiated. Each record in the dataset has 78 network features extracted with a network traffic analyser, CICFlowMeter \cite{Sharafaldin}.

We removed ten features that had no variance and seventeen highly correlated features from the dataset since they were bringing redundant information to the model. A detailed description of our pre-processing can be found on the github repository\footnotemark[\value{footnote}]. The dataset was then split into training, validation, and testing sets in the proportions of $60$\%, $20$\%, and $20$\% respectively. The CICIDS2017 dataset contains many outliers. Therefore, we used the quantile transformation to deal with outliers and transform our features to a uniform distribution [0, 1].

Table \ref{tab:class_distribution} shows how we created new attack classes by merging minority classes that had similar characteristics and behaviour. After merging the classes, we explored different class balancing techniques to further reduce the class imbalance rate and improve the prevalence ratio of the dataset. The ``Infiltration'' class was removed because of its small portion of the overall dataset.

We then performed a \ac{PCA} transformation on the CICIDS2017 dataset to reduce the dimensional feature space and simplify the learning process. As stated in the previous section, it is important to preserve as much information as possible. After a series of experiments we concluded that $99$\% of the explained variance can be retained for only 25 principal components. The feature space has been almost halved without losing any significant information.

\subsection{\ac{MLP} and \ac{DBN} Architecture}
\label{subsec:model architecture}

\begin{figure}[t]
    \centering
    \begin{subfigure}[b]{0.4\textwidth}
        \centering
        \includegraphics[width=0.6\textwidth]{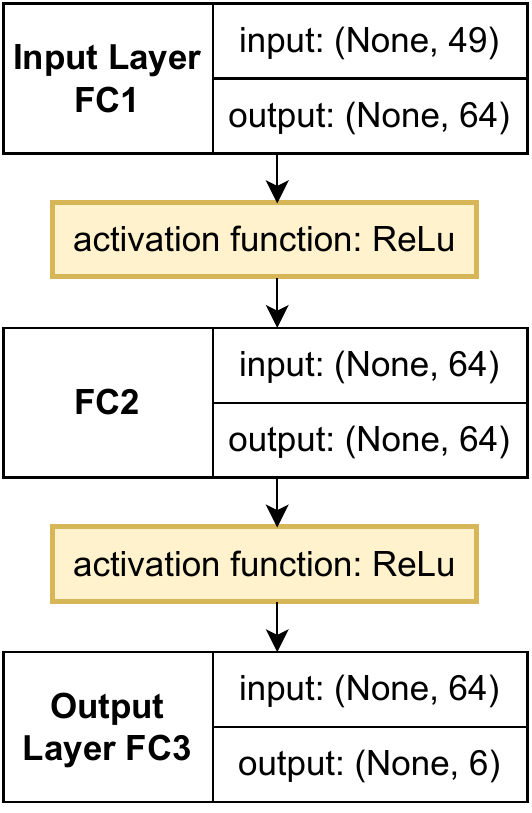}
        \caption{MLP architecture}
        \label{fig:mlp_architecture}
    \end{subfigure}
    \hfill
    \begin{subfigure}[b]{0.4\textwidth}
        \centering
        \includegraphics[width=0.6\textwidth]{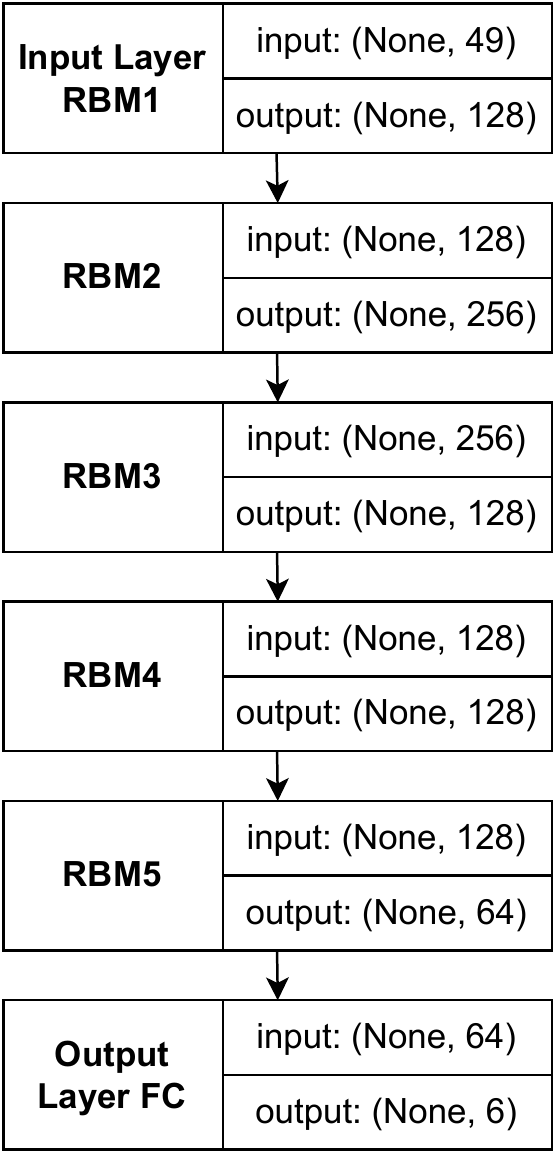}
        \caption{DBN architecture}
        \label{fig:dbn_architecture}
    \end{subfigure}
    \caption{Model architecture}
    \label{fig:model_architecture}
\end{figure}

In this study, we have implemented and tuned two different deep learning classifiers, \ac{MLP} and \ac{DBN}. Fig. \ref{fig:mlp_architecture} shows the architecture of the implemented \ac{MLP}. It consists of multiple fully connected layers. $49$ nodes were used in the input layer to represent the number of input features. After fine-tuning, two hidden layers with 64 nodes each were set and the ReLU activation function was used in the hidden layers. Six nodes were used in the output layer, each node representing one class. Finally, the Soft-Max function was used in the output layer to perform a multi-class classification.

Fig. \ref{fig:dbn_architecture} shows the architecture of the implemented \ac{DBN}. After fine-tuning, five RBMs are stacked with (49, 128), (128, 256), (256, 128), (128, 128), and (128, 64) visible/hidden nodes set per RBM respectively. The output from the last RBM is connected to a fully connected layer with 6 nodes for multi-class classification using the Softmax function. The training parameters of the \ac{DBN} and \ac{MLP} are shown in Table \ref{tab:model_design_parameters}.

\begin{table}[t]
\caption{Model design and parameters.}
\label{tab:model_design_parameters}
    \begin{subtable}[t]{0.5\textwidth}
        \centering
        \caption{DBN}
        \label{tab:dbn_parameters}
        \begin{tabular}[t]{lcc}
        \toprule
        Parameter       & Pre-training       & Fine-tuning\\
        \midrule
        Epochs          & 10                 & 30\\
        Learning rate   & 0.1                & 0.001\\
        Batch size      & 64                 & 128\\
        Momentum        & 0.9                & -\\
        Optimiser       & SGD                & Adam\\
        Loss function   & -                  & cross-entropy\\
        Gibbs step      & 1 step             & -\\
        Weight init.    & Xavier initialiser & -\\
        Bias init.      & Zeros (0)          & -\\
        \bottomrule
        \end{tabular}
    \end{subtable}
    \hfill
    \begin{subtable}[t]{0.35\textwidth}
        \centering
        \caption{MLP}
        \label{tab:mlp_parameters}
        \begin{tabular}[t]{lcc}
        \toprule
        Parameter & \\
        \midrule
        Epochs             & 10\\
        Learning rate      & 0.02\\
        Batch size         & 64\\
        Momentum           & 0.9\\
        Weight decay       & -\\
        Optimiser          & SGD\\
        Loss function      & cross-entropy\\
        \bottomrule
        \end{tabular}
    \end{subtable}
\end{table}

\subsection{Results}
\label{subsec:results}

We compare the performance of the different class balancing techniques using the evaluation metrics defined in Section \ref{subsec:model training}.

\begin{figure}[ht]
    \centering
    \includegraphics[width=0.99\textwidth]{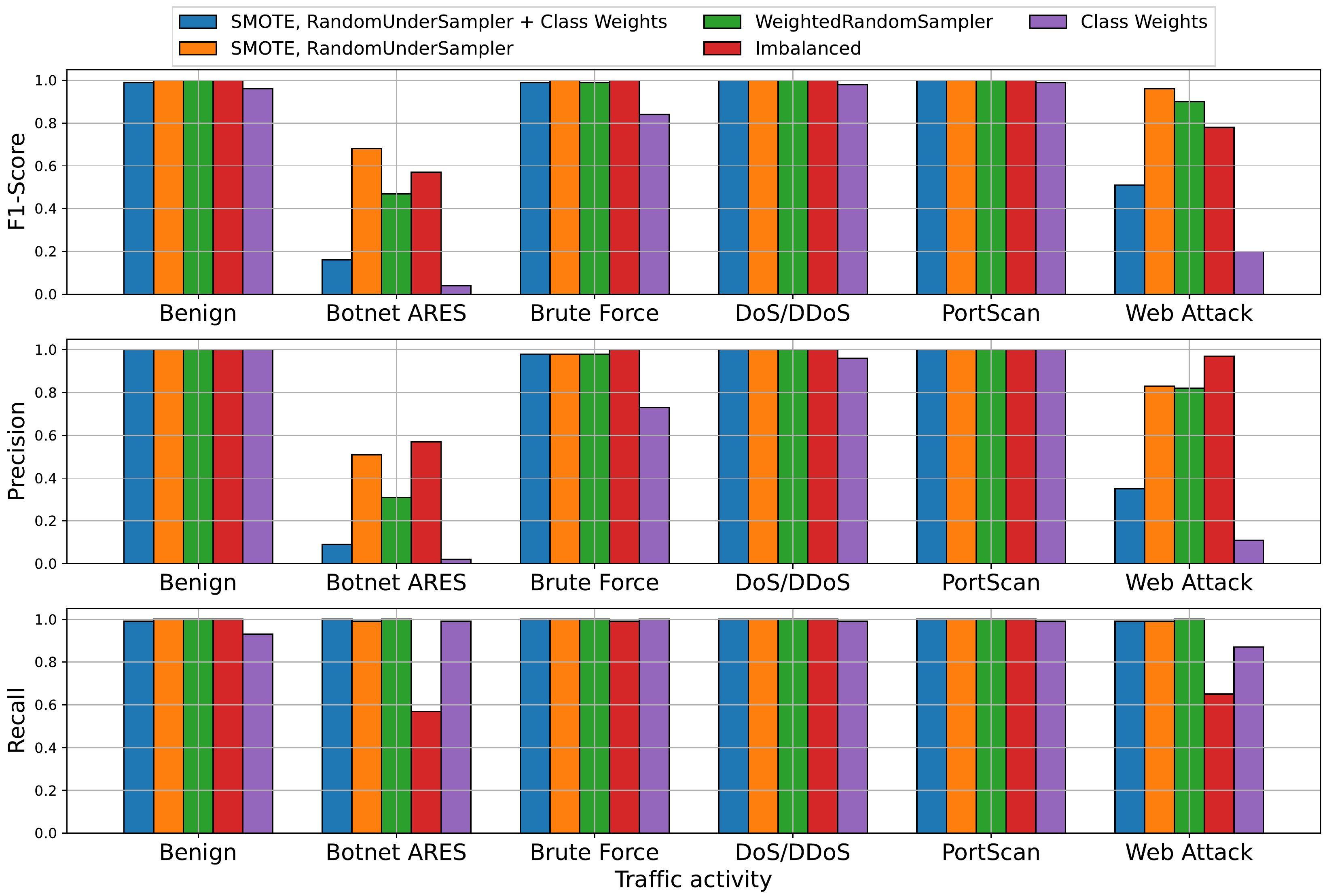}
    \caption{F1-Score, Precision and Recall achieved by the proposed \ac{DBN}-based multi-class classification for the different balancing techniques}
    \label{fig:balancing_techniques_eval}
\end{figure}

Fig. \ref{fig:balancing_techniques_eval} shows the F1-score, precision, and recall that the \ac{DBN}-based NIDS achieved for various class balancing techniques. The model has a high precision and recall, thus a high F1-score, for the ``Benigh'', ``Brute Force'', ``DoS/DDoS'', and ``PortScan'' classes. However, for the ``Web Attack'' and ``Botnet'' classes, either high recall and low precision, or vice versa is observed. It may not always be possible to have high values for both precision and recall. Hence, a trade-off between the precision and the recall is required. In the context of cyber-security, it is important to be able to detect all the malicious activities on the network (high recall). However, it is less severe if few false alarms are raised (low precision). Fig. \ref{fig:balancing_techniques_eval} shows that the combination of \ac{SMOTE} and random under-sampler achieved the best results. The recall value of every class is at least equal to $99$\%, which means that most of the attacks are successfully detected. Moreover, this method has the best precision for the Botnet class in comparison to the other class balancing techniques.

\begin{table}[ht]
    \caption{Confusion matrices.}
    \label{tab:confusion_matrix_eval}
    \begin{subtable}[ht]{\textwidth}
        \centering
        \caption{MLP}
        \label{subtab:mlp_confusion_matrix}
        \begin{tabular}[t]{lccccccc}
        \toprule
        \bf Actual & \multicolumn{6}{c}{\bf Predicted} & \bf Recall\\ 
        \cmidrule{2-7} &
        Benign & Botnet & Brute Force & DoS/DDoS & PortScan & Web Attack & \\\midrule
        Benign      & 360810  & 500  & 29   & 101    & 7      & 407  & $100$\%\\
        Botnet & 6       & 381  & 0    & 0      & 0      & 0    & $98$\% \\
        Brute Force & 4       & 0    & 1689 & 0      & 0      & 0    & $100$\%\\
        DoS/DDoS    & 113     & 0    & 0    & 63717  & 0      & 17   & $100$\%\\
        PortScan    & 3       & 4    & 1    & 23     & 11366  & 4    & $100$\%\\
        Web Attack  & 2       & 0    & 0    & 2      & 0      & 408  & $99$\%\\\midrule
        \bf Precision & $100$\% & $43$\% & $98$\% & $100$\%  & $100$\%  & $49$\% & \\
        \bottomrule
        \end{tabular}
    \end{subtable}
    \hfill
    \begin{subtable}[ht]{\textwidth}
        \centering
        \caption{DBN}
        \label{subtab:dbn_confusion_matrix}
        \begin{tabular}[t]{lccccccc}
        \toprule
        \bf Actual & \multicolumn{6}{c}{\bf Predicted} & \bf Recall\\ 
        \cmidrule{2-7} &
        Benign & Botnet & Brute Force & DoS/DDoS & PortScan & Web Attack & \\\midrule
        Benign      & 361350  & 358  & 28   & 52     & 6      & 60   & $100$\%\\
        Botnet & 3       & 384  & 0    & 0      & 0      & 0    & $99$\% \\
        Brute Force & 3       & 0    & 1691 & 0      & 0      & 0    & $100$\%\\
        DoS/DDoS    & 119     & 0    & 0    & 63707  & 0      & 21   & $100$\%\\
        PortScan    & 6       & 4    & 0    & 17     & 11371  & 3    & $100$\%\\
        Web Attack  & 5       & 0    & 0    & 1      & 0      & 406  & $99$\% \\\midrule
        \bf Precision   & $100$\%   & $51$\% & $98$\% & $100$\%  & $100$\%  & $83$\% & \\
        \bottomrule
        \end{tabular}
    \end{subtable}
\end{table}

Table \ref{tab:confusion_matrix_eval} shows the confusion matrix of the \ac{DBN} and \ac{MLP} on the testing set. The number of misclassifications and correct classification are summarised with count values for each label. We can see that both models can correctly classify most of the network traffic samples. However, significant differences can be seen between these two models in terms of precision. The \ac{MLP} classifies precisely only $49$\% of the ``Web Attack''; 407 benign traffic packets were misclassified as ``Web Attack''. On the other hand, the \ac{DBN} model classifies ``Web Attacks'' with a precision of $83$\%. This suggests that the \ac{DBN} model has developed a more meaningful pattern for these attacks, potentially due to its two-stage training.

We also compare our \ac{DBN} and \ac{MLP} models against the state-of-the-art intrusion detection approaches that use the CICIDS2017 dataset. All studied approaches trained their models using supervised learning methods. However, some of the studies perform binary rather than multi-class classification. We compare our approach against six methods. DeepGFL \cite{DeepGFL} is a framework that can extract deep features from attributed network flow graphs. LSTM and 1D-CNN \cite{1D-CNN-LSTM} are deep learning models that perform binary classification. 1D-CNN+LSTM \cite{1D-CNN-LSTM} is a combination of the two latter models for binary classification. LUCID \cite{LUCID} uses \ac{CNN}s to detect \ac{DDoS} attacks. Table \ref{tab:comparison} summarises the performance results of the aforementioned models and compares them against our models.

\begin{table}[t]
\caption{Performance comparison against existing methods using the CICIDS2017 dataset, the target number of classes is shown in the last column.}
\label{tab:comparison}
\centering
\begin{tabular}{p{2cm}p{2.5cm}cccc}
\toprule
\bf Study          & \bf Method   & \bf F1-score\ \    & \bf Recall\ \  & \bf Precision\ \  &  \bf \#Classes\\
\midrule
Our Study          & DBN          & 0.940    & 0.997    & 0.887   & 6 \\
Our Study          & MLP          & 0.873    & 0.995    & 0.817   & 6 \\
\cite{DeepGFL}     & DeepGFL      & 0.531    & 0.448    & 0.948   & 12\\
\cite{1D-CNN-LSTM} & MLP          & 0.872    & 0.862    & 0.884   & 2 \\
\cite{1D-CNN-LSTM} & LSTM         & 0.895    & 0.898    & 0.984   & 2 \\
\cite{1D-CNN-LSTM} & 1D-CNN       & 0.939    & 0.901    & 0.981   & 2 \\
\cite{1D-CNN-LSTM} & 1D-CNN+LSTM  & 0.982    & 0.991    & 0.974   & 2 \\
\cite{LUCID}       & LUCID        & 0.996    & 0.999    & 0.993   & 2 \\
\bottomrule
\end{tabular}
\end{table}

The DeepGFL \cite{DeepGFL} framework used for the classification of twelve classes produced worse results than our proposed approach in terms of recall and F1-score. The DeepGFL approach classified correctly only $44.8$\% of the malicious activities. The differences in the results could be explained either by the nature of the model itself or the different pre-processing technique used.

Furthermore, our \ac{DBN} approach achieved higher classification results than the \ac{MLP}, 1D-\ac{CNN} and LSTM in \cite{1D-CNN-LSTM}. Although these classifiers were designed for the detection of \ac{DDoS} attacks, our approach performed better while classifying six different types of attacks. For instance, the \ac{MLP} proposed by \cite{1D-CNN-LSTM} and our \ac{MLP} achieve the same F1-score but differences can be seen in terms of precision and recall. Although our \ac{MLP} presents a slightly smaller precision there has been a major improvement in the recall. As mentioned in Section \ref{subsec:results}, we prefer higher recall over precision since it is fundamental to be able to detect all malicious activities on computer networks even if it comes at the expense of a few false positives.
Finally, LUCID \cite{LUCID} and 1D-\ac{CNN}+LSTM \cite{1D-CNN-LSTM} performed slightly better than our \ac{DBN} approach. However, these two models were only used for binary classification whereas our \ac{DBN} approach achieved a high recall while performing a multi-class classification.
\section{Conclusion and Future Work}
\label{sec:conclusion}

In this paper, we researched the use of \acp{DBN} for Network Intrusion Detection. We developed two \ac{NIDS} based on \acp{DBN} and \acp{MLP}. We conducted multiple experiments using the CICIDS2017 dataset with various class-balancing techniques. The proposed \ac{MLP}-based and \ac{DBN}-based \ac{NIDS} achieved an F1-score of $87.3$\% and $94$\% respectively using a combination of \ac{SMOTE} and random under-sampler. Our experimental results demonstrate that \acp{DBN} surpass traditional \acp{MLP} on the classification of network intrusions, especially when the attacks have a small number of samples. Moreover, we compared our proposed \ac{DBN}-based \ac{NIDS} against the state-of-the-art ML-based IDS methods. Our \ac{NIDS} outperforms all other multi-class classification approaches and shows competitive results against binary classification methods with a major improvement in terms of recall. 

Currently, our method requires a centralised network sampler in order to monitor network traffic for our \ac{DBN}-based \ac{NIDS}. However, in certain applications this may not be possible due to increased data volumes, relevant privacy issues or complex network configurations. As such, a distributed approach might be necessary. We are currently working towards this direction exploring the various capabilities of \acp{DBN} when deployed in a distributed manner.

%
%
\bibliographystyle{splncs04}
\bibliography{mybibliography}

\end{document}